\documentclass[12pt,authoryear]{elsarticle}

\usepackage{amsmath,amssymb,fullpage,graphicx,tocbibind,url}

\usepackage[labelfont=bf]{caption}

\makeatletter
\def\ps@pprintTitle{%
\let\@oddhead\@empty
\let\@evenhead\@empty
\def\@oddfoot{\footnotesize\itshape\hfill\today}%
\let\@evenfoot\@oddfoot}
\makeatother

\begin{document}


\title{\textbf{New economic windows on income and wealth: The $\boldsymbol{\kappa}$-generalized family of distributions\tnoteref{t1}}}

\tnotetext[t1]{We wish to thank Giorgio Kaniadakis, the father of the $\kappa$-distribution, and Paul Ormerod, who encouraged us to write this article. We are also grateful to (in random order): Simone Landini, Costantino Tsallis, Jean-Paul Fitoussi and Joe Stiglitz. Endless discussions with our group in Ancona, in particular Ruggero Grilli, Antonio Palestrini, Luca Riccetti and Alberto Russo, have been very stimulating. We thank all of them for their constructive and interesting insights shared during the meetings in which this article was discussed.}

\author{F.~Clementi\corref{cor1}}
\ead{fabio.clementi@unimc.it}
\address{University of Macerata, Department of Political Science, Communication and International Relations, Piazza San Vincenzo Maria Strambi 1, 62100 Macerata, Italy}

\author{M.~Gallegati\corref{}}
\ead{mauro.gallegati@univpm.it}
\address{Polytechnic University of Marche, Department of Management, Piazzale Raffaele Martelli 8, 60121 Ancona, Italy}

\cortext[cor1]{Corresponding author}

\date{\today}


\begin{abstract}
Over the last decades, the distribution of income and wealth has been deteriorating in many countries, leading to increased inequalities within and between societies. This tendency has revived the interest in the subject greatly, yet it still receives very little attention within the realm of mainstream economic thinking. One reason for this is that the basic paradigm of ``standard economics'', the representative-agent General Equilibrium framework, is badly equipped to cope with distributional issues. Here we argue that when the economy is treated as a complex system composed of many heterogeneous interacting agents who give rise to emergent phenomena, to address the main stylized facts of income/wealth distribution requires leaving the toolbox of mainstream economics in favour of alternative approaches. The ``$\kappa$-generalized'' family of income/wealth distributions, building on the categories of complexity, is an example of how advances in the field can be achieved within new interdisciplinary research contexts.
\end{abstract}

\maketitle


\section{Introduction}

The renewed interest in the problems of income (and wealth) distribution is exemplified by recent work of contemporary prominent economists that followed the resurgence of inequality after the global crisis of 2008--09 \citep{Stiglitz2012,Piketty2014,Atkinson2015,Stiglitz2015}. In the developed world, the immediate effect of the economic crisis was less inequality: 2008 and 2009 saw a decline in the percentage of income retained by the rich mostly due to capital losses suffered by people in this category. The compressed income distribution immediately following the crisis was cancelled out in the years since 2010. The economic recovery, although fragile, has been most beneficial for the well-off: it had a powerful and immediate effect on capital income, whereas the impact on unemployment\textemdash and consequentially the overall economy\textemdash took years to materialize, at which point capital income had the time to rebound back from being bent.

The rise of inequality immediately after the economic crisis is not a new phenomenon. The end of the 1970s was the point of departure for widening disparities of income and wealth in most developed countries, marking the end of a half-century of declining or relatively stable inequality. In emerging economies, economic growth has helped to reduce sharply the prevalence of poverty, but at the same time high levels of inequality have risen further.

The growth of inequality reignited by the crisis is therefore part of a longer trend that began at least two decades prior to the onset of the crisis itself, but due to the crisis it has now become a universal concern among both policy makers and societies at large. The social compact has indeed started to show signs of unravelling in many countries: believing that they are bearing the brunt of a crisis for which they have no responsibility, while people at the top of the distribution appear to have been spared, uncertainty and inequality-related issues have reached the lower and middle classes in many societies, thus adding urgency to deal with policy issues related to inequality.

The crisis that we have gone through has triggered research on the role that rising inequality has played in creating preconditions for the collapse, with many recent studies suggesting that deteriorating distribution tends to amplify the risk of macroeconomic instability \citep{Rajan2010,OstryBerg2011,Cingano2014,OstryBergTsangarides2014,DablaNorrisKochharSuphaphiphatRickaTsounta2015,KumhofRanciereWinant2015}. But the economic crisis has also persuaded many scholars that the time has come for a serious rethinking of standard economic theory. The reasons for this are due to fundamental problems with models based on the General Equilibrium approach, which are both unsound theoretically and incompatible with the data \citep{Kirman2010}. In particular, the unrealistic individual basis for aggregate behaviour\textemdash the so-called ``representative-agent'' framework\textemdash imposes a straitjacket on the mainstream box of tools that inhibits it from any application on distribution \citep{ClementiGallegati2016}. As we will see, we should rather analyse the economy as a complex adaptive system and take the network structure that governs interaction among heterogeneous agents into account to give distributional issues the conceptual and methodological attention they merit from the economics profession.


\section{A new way of thinking about the income and wealth distribution problem}

For so long, the subject of personal income/wealth distribution has been playing a peripheral role in economic analysis \citep{Atkinson1997}. Both classical economists and mainstream economic theory were certainly concerned with the determinants of payments to factors\textemdash labour, land and capital\textemdash but the relationship of the functional distribution with the distribution by size was typically not spelled out. In particular, the dominant paradigm in macroeconomics has been criticized for having long ignored (re-)distributional issues by reasoning within a modelling framework in which all distributional considerations are assumed away\textemdash the so-called ``representative-agent'' framework.

In representative-agent models, which have become ubiquitous in economic teaching and research since the 1970s, the aggregate behaves as if it were the result of a ``rational'' individual's decisions \citep{Hartley1997}. But to take this route means assuming from the outset that all agents have identical preferences and resources\textemdash i.e. receive the same wage, are endowed with the same wealth and enjoy the same sources of income\textemdash so that the issue of inequality in the distribution of wealth and income is totally avoided \citep{Piketty2014}. Probably, many macro-economists would justify the use of a representative-agent model on the ground that, while it is not precisely true that all the agents have identical endowments and tastes, it is nevertheless a close enough approximation. But in this case being ``close'' does not count: in order to argue that the aggregate behaviour of a large system of individuals is typically that of an average or representative member of the population, one must believe that is precisely true that all agents have the same endowments and preferences \citep{KirmanKoch1986,Kirman1992}. Similarly, representative-agent models cannot address the consequences of changes to the distribution of income and wealth, and so they cannot address the impact of redistributive policies on inequality\footnote{A representative agent has welfare significance only if lump-sum redistribution of endowments is possible \citep{AtkinsonStiglitz2015}.} or the impact of inequality on macroeconomic variables which is centre stage in recent economic debate \citep{StiglitzOcampoSpiegelFfrenchDavisNayyar2006}.

How can this be changed? It is suggested to drop the unrealistic individual basis for aggregate behaviour and the even more unreasonable assumption that the aggregate behaves like a ``rational individual''. The economy should be rather regarded as a complex system composed of many heterogeneous agents whose actions cannot be considered in isolation and only linked through an anonymous market. Their behaviour is constantly and mutually influenced by others. When agents are directly linked to each other and influence each other, the relationship between the behaviour of individuals and the behaviour of aggregate variables will be different from that in the anonymous market situation, in which all agents are linked to each other only through the price system. What we observe at the aggregate level will not mimic what we observe at the individual level, nor will it correspond to the behaviour of some ``representative individual''. The outcome of this process of interaction is not just the aggregation of individual behaviour, but far richer because the system is complex \citep{Anderson1972}. Some aspects of complex systems spontaneously emerge as the result of long-term endogenous build-up, like the emergence of income and wealth distributions with Paretian (power-law) tails.\footnote{Statistical physics has also revealed that the laws of power, to which class belongs the Pareto's law\textemdash the observation that where a large number of agents contribute to a result, the majority of the result is due to the contributions of a minority of agents\textemdash, are the signature of ``self-organized criticality'' \citep{Bak1996}, i.e. the mechanism opposed to ``top-down'' General Equilibrium economics.}

The interaction among agents can be identified with the \textit{causa causans} that originates the distribution as an emergent phenomenon. As a result of their interactions, heterogeneous agents can change their status, \textit{ergo} the way they are distributed on different possible states of the system can mutate. This implies that each state of the system will be otherwise populated at any given time but also at different times. When appropriate conditions occur that insist on the system and affect its constituents, one can reach a stationary configuration where the number of agents in different states does not change any more but individual agents may continue to change their status. At the same time, there is an individual disequilibrium and aggregate balance.

The observed distribution of income/wealth can represent a balance of the system, although its constituents are not because they continue to change status (e.g. some rich become poor, and vice versa). Exchanges in this case are compensatory: if one moves from B to enter the state A ($\text{rich}\rightarrow\text{poor}$), then one comes from the state B to enter A ($\text{poor}\rightarrow\text{rich}$). If this mechanism is indefinitely repeated in time, then we are facing a situation that, by analogy with the economic jargon but with little accuracy, one may define  as a \textit{steady state}, i.e.: the distribution was carried to the stationary state because the total number of agents in all possible states of the system does not change even though individual agents continue to change status. This is the basic idea of \textit{statistical equilibrium} \citep{Foley1994,Aoki1996,Aoki2002,AokiYoshikawa2007}.

In this context, one cannot use the representative-agent framework, or even a variation of it, because it has many logical contradictions and is completely unfitted to describe a complex system. Economic systems are complex because their constituents (agents) are characterized by the categories of \textit{heterogeneity} and \textit{interaction}, which feed off each other and overlap to originate \textit{emergent phenomena}. As a result, the aggregate behaviour is very different from what would be the behaviour of a single individual: this is the result of a mechanism which is described microscopically by functions only exceptionally linear, and the sum of the functions of the parts is not the aggregate function \citep{LandiniGallegati2014}.

Rather, the complexity typical of the economic system requires a statistical equilibrium approach: the possibility of equilibrium at the macro level in the presence of a multitude of situations of non-equilibrium at the microscopic level. Statistical equilibrium is one of the basic concepts of the statistical mechanics approach to the modelling of complex systems, which directly connect the relevant microscopic information to useful macroscopic quantities by means of probabilistic rules. The large number of interacting particles (agents) does not make it possible, indeed, to proceed with the deterministic method of classical mechanics \citep{Khinchin1948}. Furthermore, economic systems constitute of agents which are not all the same as atoms of an ideal gas but heterogeneous with regard to both endowments and behaviour. The continuous overlapping between heterogeneity and interaction generates income/wealth inequality and the associated distribution shape.

When income/wealth flows/accumulates to people in a particular class, all members of this class are ``identical'' with respect to the class they are enrolled because of their level of income/wealth (\textit{weak heterogeneity}) but not as compared to the way income or wealth enabling access to different classes has been gained (\textit{strong heterogeneity}). Therefore, disparities cannot be explained only by relative differences in levels of income and wealth (weak heterogeneity); they also result from the way agents' action has been allowed to take steps toward those levels (strong heterogeneity). Put differently, in addition to talent and effort, which determine relative income/wealth positions, unequal opportunities\textemdash economic, political and social\textemdash that prevent agents to reach their full potential can also be held responsible for unequal outcomes.

Furthermore, as in physical collisions particles with very large energy tend to exchange only a small part of their energy when colliding with less energetic particles, in economic interactions the richest individuals tend to put at stake, in their interactions, only a small part of their income/wealth, because of their higher propensity to save. This enables them to protect their status and exclude the rest from joining their ``club'', leading to reinforcement of heterogeneity.

These considerations lead to an ``exclusion'' principle that calls for different probabilistic laws (\textit{relativism}) depending on the part of the income/wealth support the analysis is focused on \citep{ClementiGallegati2016}: as in statistical mechanics many bosons obey the Bose-Einstein statistics at low energies and give rise to the family of exponential distributions, whereas few fermions obey the Fermi-Dirac statistics at high energies and follow a power law, at the bottom of the income (wealth) distribution many individuals earn (accumulate) little of it as compared to the gains of few individuals located at the top.

At the bottom of the income/wealth distribution no exclusion principle is at work: there may be as many poor (bosons) as you can have both at the time of the Pharaohs and today, with the only limit being the total number of units and total income or wealth. At the top, however, the exclusion principle is in force and becomes stronger the higher the social ladder is climbed, so that you can have a few CEOs (the strongest fermions) but only one Pharaoh (the extreme fermion). Whether obeying or not an exclusion principle determines the shape of the income/wealth distribution and represents a form of relativism in such a way that the reference probabilistic principle is the family of exponential distributions for the bottom of its support, whilst for the top the reference probabilistic principle is the family of power-law distributions.

Valid and promising tools exist that can be expected to tackle the theoretical challenges on the size distribution of income and wealth raised above. As will be shown in the following, among the many parametric models for the size distribution of income and wealth proposed in the literature, the $\kappa$-generalized is the only one which formally and explicitly embeds both the exponential and the power-law shapes in a single functional form. Therefore, when operating on the portion of the income/wealth distribution ruled by probabilistic principles of the exponential family (no exclusion principle), the $\kappa$-generalized model replicates remarkably well the underlying data, and the same good performance holds when it operates on that portion of the income or wealth support ruled by probabilistic principles of the power-law family (exclusion principle), thus capturing both the mentioned relativistic aspects.


\section{$\kappa$-generalized models of income and wealth distribution}

The interest in finding parametric models for the size distributions of income and wealth has a long history. A natural starting point in this area of inquiry was the observation that the number of persons in a population whose incomes exceed $x$ is often well approximated by $Cx^{-\alpha}$, for some real $C$ and positive $\alpha$, as Pareto argued over 100 years ago \citep{Pareto1895,Pareto1896,Pareto1897a,Pareto1897b}. Since the early studies of Pareto, numerous empirical works have shown that the power-law tail is a ubiquitous feature of income and wealth distributions. However, even 100 years after Pareto observation, the understanding of the shape of income/wealth distribution is still far to be complete and definitive. This reflects the fact that there are two distributions, one for the rich, following the Pareto power law, and one for the vast majority of people, which appears to be governed by a completely different law.

Over the years, research in the field has considered a wide variety of functional forms as possible models for the size distribution of income and wealth, some of which aim at providing a unified framework for the description of real-world data\textemdash including the heavy tails present in empirical income and wealth distributions \citep{KleiberKotz2003}. Among these, the ``$\kappa$-generalized distribution'' was found to work remarkably well \citep{ClementiGallegatiKaniadakis2007,ClementiDiMatteoGallegatiKaniadakis2008,ClementiGallegatiKaniadakis2009,ClementiGallegatiKaniadakis2010,ClementiGallegatiKaniadakis2012a,ClementiGallegati2016}. First proposed in 2007, and further developed over successive years, this model finds its roots in the context of generalized statistical mechanics \citep{Kaniadakis2001,Kaniadakis2002,Kaniadakis2005,Kaniadakis2009a,Kaniadakis2009b,Kaniadakis2013}. Within this theoretical framework, the ordinary exponential function $\exp\left(x\right)$ generalizes into the function $\exp_{\kappa}\left(x\right)$ defined through
\begin{equation}
\exp_{\kappa}\left(x\right)=\left(\sqrt{1+\kappa^{2} x^{2}}+\kappa x\right)^{\frac{1}{\kappa}},\quad x\in\mathbb{R},\quad\kappa\left[0,1\right).
\label{eq:Equation_1}
\end{equation}
We recall briefly that in the $\kappa\rightarrow0$ limit the function \eqref{eq:Equation_1} reduces to the ordinary exponential, i.e. $\exp_{0}\left(x\right)=\exp\left(x\right)$, and for $x\rightarrow0$\textemdash independently on the value of $\kappa$\textemdash behaves very similarly with the ordinary exponential. On the other hand, the most interesting property of $\exp_{\kappa}\left(x\right)$ for modelling the size distribution of income and wealth is the power-law asymptotic behaviour
\begin{equation}
\exp_{\kappa}\left(x\right)\underset{x\rightarrow\pm\infty}{\sim}\left|2\kappa x\right|^{\pm\frac{1}{\left|\kappa\right|}}.
\label{eq:Equation_2}
\end{equation}

Given \eqref{eq:Equation_1}, the $\kappa$-generalized distribution is defined in terms of the following cumulative distribution function (CDF)
\begin{equation}
F\left(x;\alpha,\beta,\kappa\right)=1-\exp_{\kappa}\left[-\left(\frac{x}{\beta}\right)^{\alpha}\right],\quad x>0,\quad \alpha,\beta>0,\quad \kappa\in\left[0,1\right),
\label{eq:Equation_3}
\end{equation}
where $\left\{\alpha,\beta,\kappa\right\}$ are parameters. The corresponding probability density function (PDF) reads as
\begin{equation}
f\left(x;\alpha,\beta,\kappa\right)=\frac{\alpha}{\beta}\left(\frac{x}{\beta}\right)^{\alpha-1}\frac{\exp_{\kappa}\left[-\left(\frac{x}{\beta}\right)^{\alpha}\right]}{\sqrt{1+\kappa^{2}\left(\frac{x}{\beta}\right)^{2\alpha}}}.
\label{eq:Equation_4}
\end{equation}

The distribution defined through \eqref{eq:Equation_3} and \eqref{eq:Equation_4} can be viewed as a generalization of the Weibull distribution, which recovers in the $\kappa\rightarrow0$ limit. Consequently, the exponential law is also a special limiting case of the $\kappa$-generalized distribution, since it is a special case of the Weibull with $\alpha=1$. For $x\rightarrow0^{+}$, the $\kappa$-generalized behaves similarly to the Weibull distribution, whereas for large $x$ it presents a Pareto power-law tail, hence satisfying the weak Pareto law \citep{Mandelbrot1960}.

\begin{figure}[!t]
\centering
\includegraphics[width=\textwidth]{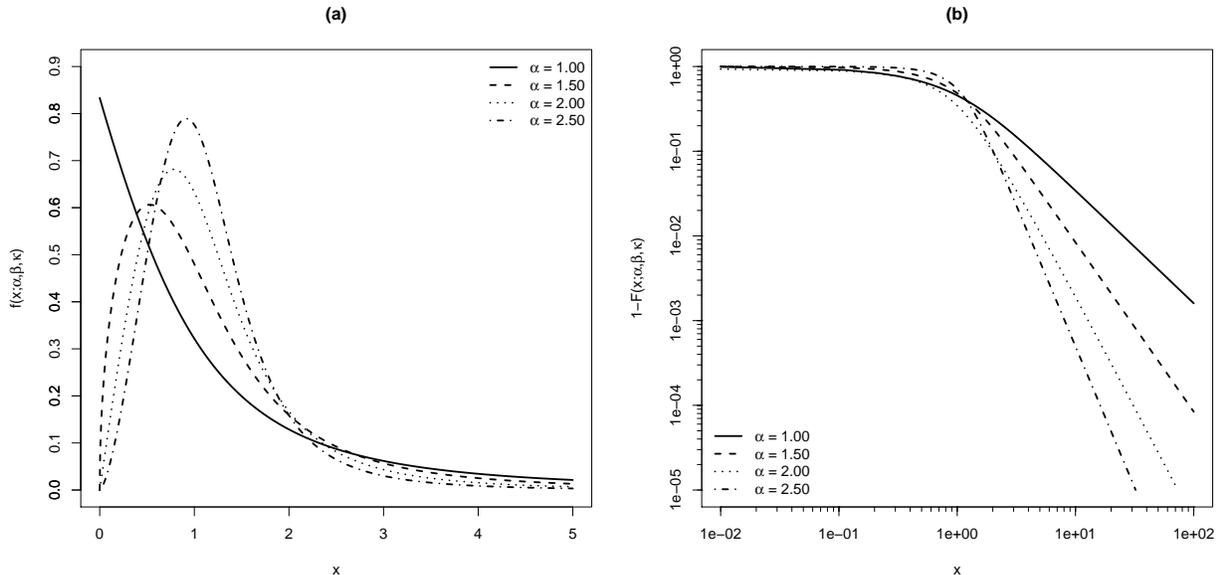}
\caption{Plot of the $\kappa$-generalized PDF (a) and CCDF (b) for some different values of $\alpha$ ($=1.00,1.50,2.00,2.50$) and fixed $\beta$ ($=1.20$) and $\kappa$ ($=0.75$). The CCDF is plotted on doubly-logarithmic axes, which is the standard way of emphasizing the right-tail behavior of a distribution. Notice that the curvature (shape) of the distribution becomes less (more) pronounced when the value of $\alpha$ decreases (increases). The case $\alpha=1.00$ corresponds to the standard exponential distribution.}
\label{fig:Figure_1}
\end{figure}

\begin{figure}[!t]
\centering
\includegraphics[width=\textwidth]{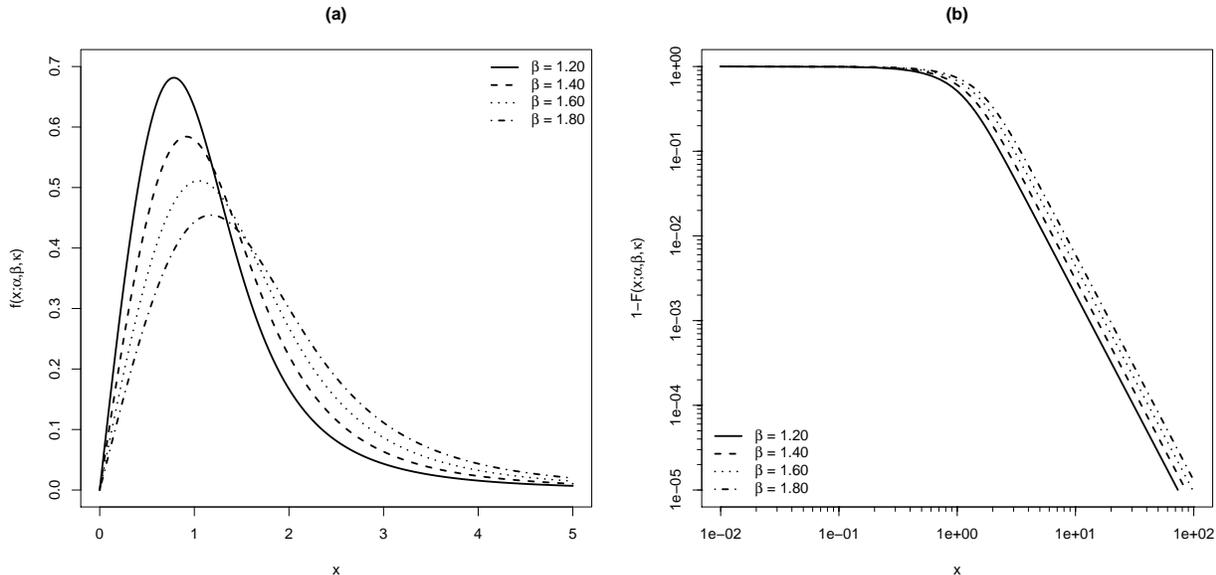}
\caption{Plot of the $\kappa$-generalized PDF (a) and CCDF (b) for some different values of $\beta$ ($=1.20,1.40,1.60,1.80$) and fixed $\alpha$ ($=2.00$) and $\kappa$ ($=0.75$). The CCDF is plotted on doubly-logarithmic axes, which is the standard way of emphasizing the right-tail behavior of a distribution. Notice that the distribution spreads out (concentrates) as the value of $\beta$ increases (decreases).}
\label{fig:Figure_2}
\end{figure}

\begin{figure}[!t]
\centering
\includegraphics[width=\textwidth]{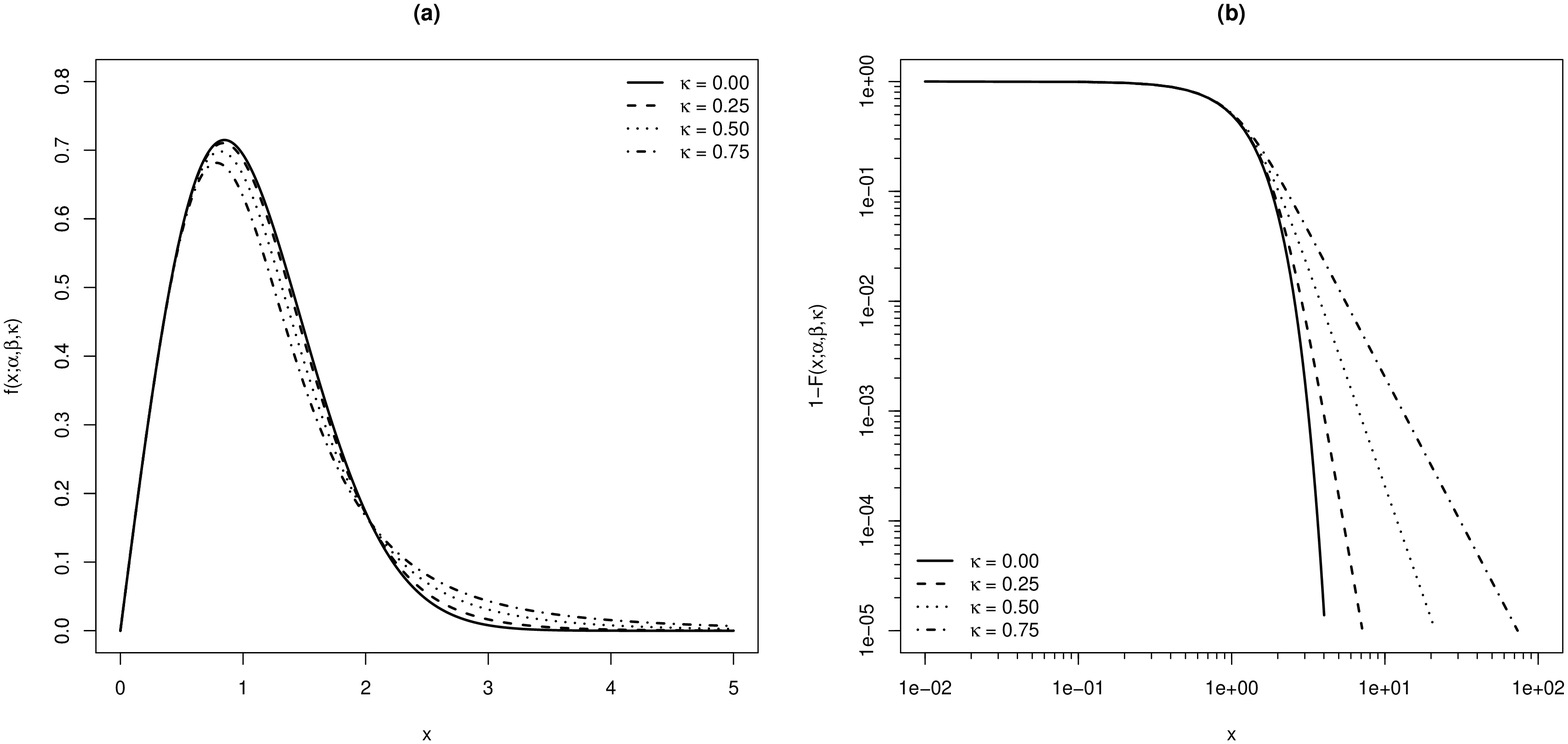}
\caption{Plot of the $\kappa$-generalized PDF (a) and CCDF (b) for some different values of $\kappa$ ($=0.00,0.25,0.50,0.75$) and fixed $\alpha$ ($=2.00$) and $\beta$ ($=1.20$). The CCDF is plotted on doubly-logarithmic axes, which is the standard way of emphasizing the right-tail behavior of a distribution. Notice that the upper tail of the distribution fattens (thins) as the value of $\kappa$ increases (decreases). The case $\kappa=0.00$ corresponds to the Weibull (stretched exponential) distribution.}
\label{fig:Figure_3}
\end{figure}

Figures \ref{fig:Figure_1} to \ref{fig:Figure_3} illustrate the behaviour of the $\kappa$-generalized PDF and complementary CDF, $1-F\left(x;\alpha,\beta,\kappa\right)$, for various parameter values. The exponent $\alpha$ quantifies the curvature (shape) of the distribution, which is less (more) pronounced for lower (higher) values of the parameter, as seen in Figure \ref{fig:Figure_1}.\footnote{It should be noted that for $\alpha=1$ the density exhibits a pole at the origin, whereas for $\alpha>1$ there exists an interior mode.}
%
%
The constant $\beta$ is a characteristic scale, since its value determines the scale of the probability distribution: if $\beta$ is small, then the distribution will be more concentrated around the mode; if $\beta$ is large, then it will be more spread out (Figure \ref{fig:Figure_2}).
%
%
Finally, as Figure \ref{fig:Figure_3} shows, the parameter $\kappa$ measures the fatness of the upper tail: the larger (smaller) its magnitude, the fatter (thinner) the tail.
%

Expressions that facilitate the analysis of the associated moments and various tools for the measurement of inequality have been reported for the $\kappa$-generalized distribution \citep{ClementiDiMatteoGallegatiKaniadakis2008,ClementiGallegatiKaniadakis2009,ClementiGallegatiKaniadakis2010,ClementiGallegatiKaniadakis2012a,ClementiGallegati2016}. These expressions are functions of the parameters in the model and prove useful in the analysis of population characteristics.

The $\kappa$-generalized distribution was also successfully used in a three-component mixture model for analysing the singularities of survey data on \textit{net} wealth, i.e. the value of gross wealth minus total debt, which present highly significant frequencies of households or individuals with null and/or negative wealth \citep{ClementiGallegatiKaniadakis2012b,ClementiGallegati2016}. The support of the $\kappa$-generalized mixture model for net wealth distribution is the real line $\mathbb{R}=\left(-\infty,\infty\right)$, thus allowing to describe the subset of economic units with nil and negative net worth. Furthermore, four-parameter variants exist that contain as a particular case the $\kappa$-generalized model for income distribution \citep{Okamoto2013}.

During the last decade, there have been several applications of $\kappa$-generalized models to real-world data on income and wealth distribution. Of special interest are papers fitting several distributions to the same data, with an eye on relative performance. From comparative studies such as \citet{ClementiGallegatiKaniadakis2010}, who considered the distribution of household income in Italy for the years 1989 to 2006 , it emerges that model \eqref{eq:Equation_4} typically outperforms its three-parameter competitors such as the Singh-Maddala \citep{SinghMaddala1976} and Dagum type I \citep{Dagum1977} distributions, apart from the GB2 which has an extra parameter.\footnote{The GB2 is a quite general family of parametric models for the size distribution of income that nests most of the functional forms previously considered in the size distributions literature as special or limiting cases \citep{McDonald1984}. In particular, both the Singh-Maddala and Dagum type I distributions are special cases of the GB2.} The model was also fitted by \citet{ClementiGallegatiKaniadakis2012a} to data from other household budget surveys, namely Germany 1984--2007, Great Britain 1991--2004, and the United States 1980--2005. In a remarkable number of cases, the distribution of household income follows the $\kappa$-generalized more closely than the Singh-Maddala and Dagum type I. In particular, the fit is statistically superior in the right tail of data with respect to the other competitors in many instances. Another example of comparative study is \citet{Okamoto2012}, who considered US and Italian income data for the 2000s. He found the three-parameter $\kappa$-generalized model to yield better estimates of income inequality even when the goodness-of-fit is inferior to that of distributions in the GB2 family. The excellent fit of the $\kappa$-generalized distribution and its ability in providing relatively more accurate estimation of income inequality have recently been confirmed in a book by \citet{ClementiGallegati2016}, who utilize household income data for 45 countries selected from the most recent waves of the LIS Database (\url{http://www.lisdatacenter.org/}).

The previously mentioned works were mainly concerned with the distribution of household incomes. In an interesting contribution by \citet{ClementiGallegatiKaniadakis2012b}, the $\kappa$-generalized distribution was used in a three-component mixture to model the US net wealth data for 1984--2011. Both graphical procedures and statistical methods indicate an overall good approximation of the data. The authors also highlight the relative merits of their specification with respect to finite mixture models based upon the Singh-Maddala and Dagum type I distributions for the positive values of net wealth. Similar results were recently obtained by \citet{ClementiGallegati2016} when analysing net wealth data for 9 countries selected from the most recent waves of the LWS Database (\url{http://www.lisdatacenter.org/}).

Finally, four-parameter extensions of the $\kappa$-generalized distribution were used by \citet{Okamoto2013} to analyse household income/consumption data for approximately 20 countries selected from Waves IV to VI of the LIS Database. To provide a comparison with alternative four-parameter models of income distribution, the GB2 and the double Pareto-lognormal (dPlN) distribution introduced by \citet{ReedJorgensen2004} were also fitted to the same datasets. In almost all cases, the new variants of the $\kappa$-generalized distribution outperform the other four-parameter models for both the income and consumption variables. In particular, they show an empirical tendency to estimate inequality indices more accurately than they counterparts do.


\section{Concluding remarks}
\label{sec:ConcludingRemarks}

For much of the past century, the subject of personal income/wealth distribution was very much marginalized by the economics profession. There are signs that in recent years it is being welcomed back, but the standard economic model has to be changed if we are to give further impetus to the re-incorporation of distributional issues into the main body of economic analysis. New theoretical perspectives and tools exist that help to address the main stylized facts of the distribution of income and wealth among individuals\textemdash e.g. why the tails of the distribution are Pareto (fat-tailed) and why at lower levels of income/wealth the distributions seem to be described by a different law. Agent-based computational economics\textemdash i.e. the computational study of economies modelled as evolving systems of autonomous interacting agents\textemdash would represent a further possibility to proceed along this field of research by allowing the definition of a theoretical model able to demonstrate the emergence of $\kappa$-generalized income/wealth distributions as the result of decentralized interactions of a large number of heterogeneous agents. The story is just beginning to unfold.


\bibliographystyle{elsarticle-harv}
\bibliography{new}


\end{document}